\documentclass{article}
\usepackage{arxiv}
\usepackage{graphicx}
\usepackage[utf8]{inputenc} 
\usepackage[T1]{fontenc}    
\usepackage{hyperref}       
\usepackage{url}            
\usepackage{booktabs}       
\usepackage{amsfonts}       
\usepackage{breqn}
\usepackage{nicefrac}       
\usepackage{microtype}      
\usepackage{lipsum}

\title{The Sci-hub Effect: Sci-hub downloads lead to more article citations}

\author{
  Juan C. Correa\thanks{\url{https://correajc.com/}. Department of Management, Faculty of Business Administration, University of Economics, Prague, Czech Republic} \\
  Faculty of Business Administration\\
  University of Economics, Prague, Czechia\\
  \texttt{juan.correa@vse.cz} \\
   \And
 Henry Laverde-Rojas \\
  Faculty of Economics\\ 
  Universidad Santo Tomás, Bogotá, Colombia\\
  \texttt{henrylaverde@usantotomas.edu.co} \\
   \And
   Fernando Marmolejo-Ramos \\
   Centre for Change and Complexity in Learning \\
   University of South Australia \\
   \texttt{Fernando.Marmolejo-Ramos@unisa.edu.au} \\
   \And
   Julian Tejada \\
   Departamento de Psicologia \\
   Universidade Federal de Sergipe \\
   \texttt{julian.tejada@gmail.com} \\
   \And
   Štěpán Bahník \\
   Faculty of Business Administration \\
   University of Economics, Prague, Czechia \\
   \texttt{stepan.bahnik@vse.cz} \\
}

\begin{document}
\maketitle

\begin{abstract}
Citations are often used as a metric of the impact of scientific publications. Here, we examine how the number of downloads from Sci-hub as well as various characteristics of publications and their authors predicts future citations. Using data from 12 leading journals in economics, consumer research, neuroscience, and multidisciplinary research, we found that articles downloaded from Sci-hub were cited 1.72 times more than papers not downloaded from Sci-hub and that the number of downloads from Sci-hub was a robust predictor of future citations. Among other characteristics of publications, the number of figures in a manuscript consistently predicts its future citations. The results suggest that limited access to publications may limit some scientific research from achieving its full impact.
\end{abstract}

\keywords{Sci-hub \and Citations \and Scientific Impact \and Scholar Consumption \and Knowledge dissemination}

\section*{Introduction}
Science and its outputs are essential in daily life, as they help to understand our world and provide a basis for better decisions. Although scientific findings are often cited in social media and shared outside the scientific community \cite{Milkman2014}, their primary use is what we could call ``scholar consumption.'' This phenomenon includes using websites that provide subscription-based access to massive databases of scientific research \cite{Boukacem2016}. Despite the existence of local databases in each region of the world, \textit{Web of Science} and \textit{Scopus} are the most widely employed \cite{Zhu2020}. The use of these databases, however, is changing given the emergence of alternative services such as Library Genesis, Paperhub, or Sci-hub, which provide free access to scientific publications without regard for their copyright status \cite{Seguin2019}. Sci-hub, in particular, has gained worldwide renown as an initiative which poses wide-ranging implications for all belonging to the global academic system \cite{Mejia2017, Bendezu2016,Strielkowski2017, Himmelstein2018,Nicholas2019}.

As Sci-hub gives access to nearly all scientific literature to anyone with an internet connection, some scholars suggest that the traditional business model that relies on subscription to journals may become unsustainable \cite{mcnutt_my_2016,Himmelstein2018,Strielkowski2017}. Access to scientific literature has been unequal between countries of different levels of economic development \cite{Horowitz1986}. Sci-hub allows access to scientific literature for scientists of both developed and developing countries, and it may, therefore, decrease the difference in access to scientific publications between nations. Because the collective productive knowledge provided by scientific literature might be one of the driving mechanisms of economic development \cite{Hausmann2013,Jaffe2013, Laverde2019, Pinto2020}, more access to this literature would translate into more prosperity, especially for the countries which used to have limited access.

A more immediate impact of Sci-hub is likely to be observed on scientific knowledge interchange in general, and scientific citing in particular. Existing studies have examined various other predictors of citations. The length of the title of a paper has been shown to be negatively associated with its annual citation rate \cite{Lee2014}. According to some scholars, the use of graphs and tables for communicating scientific findings could be critical as well \cite{Smith2002}. The area of knowledge and the impact factor of journals have also been deemed as relevant predictors of citations  \cite{paulus2015,varki2017,zhang2017}. Moreover, the number of citations is influenced by the number of authors per article \cite{Breitzman2015}  and the so-called ``\textit{chaperone effect in scientific publishing}'' that takes place when senior scientists appear as the last author of a paper whose first author is a junior scientist \cite{Sekara2018}. Despite previous studies suggesting that research articles downloads can predict its later citations \cite{Brody2006}, up to the best of our knowledge, the role of Sci-hub as a relevant predictor of citations remains unknown.

Our goal in this paper is to study the effect of Sci-hub on citing published articles. Our central hypothesis is that the access of articles through Sci-hub is associated with a higher number of future citations because it bypasses the obstacles imposed by paid subscription-based information retrieval services, and thus it increases the potential impact of the article \cite{Himmelstein2018}. Additionally, we test the effects of various other characteristics of the article, the journal where it is published, and its authors in predicting its citation rate. The generalized specification of our model is as follows:

\begin{equation}
C_i = \beta \times Scihub_i + X_{i}^{'} \gamma + \theta_i
\label{eq1}
\end{equation}

Where $C_i$ stands for the number of citations the paper $i$ has received, as captured by Scopus official records. $\beta$ is our parameter of interest which quantifies the relationship between the citation of a paper and the number of times the paper $i$ was downloaded from Sci-hub. $X_{i}'$ is a vector containing the following control variables: i) the impact factor of the journal where the paper was published, ii) the length of the title of the paper, as captured by the number of types or unique words in it, iii) the number of figures included in the paper $i$ and its supplementary material, iv) the number of tables included in the paper $i$ and its supplementary material, v)  The chaperone effect, as captured by the H-index of the first and last author of the paper $i$ and the number of publications of the last author in the same journal, in case that the last author had a bigger H-index than the first author, and vi) the annual GDP per capita and the Nature Index for country of affiliation of the first and the last author. The annual GDP per capita was obtained from World Bank, and the Nature Index provides a close to real-time proxy of high-quality research output and collaboration at the institutional, national, and regional level. The Nature index in our analyses is used only as a national proxy. Finally, the parameter $\theta_i$ represents the residuals of our model.

\section*{Materials and method}
\textbf{Data.} 
To estimate the empirical impact of Sci-Hub on the citations of papers, we used two data sets containing information about articles published in twelve selected journals. Three of these journals publish multidisciplinary research (``\textit{Nature}'', ``\textit{Science}'', and ``\textit{Proceedings of the National Academy of Sciences}''), and the remaining were specialist journals in economics (``\textit{The Quarterly Journal of Economics}'', ``\textit{Journal of Political Economy}'', and ``\textit{Econometrica}), consumer research (``\textit{Journal of Consumer Research}'', ``\textit{Journal of Retailing and Consumer Services}'', and ``\textit{Journal of Consumer Psychology}''), and neuroscience (``\textit{Nature Reviews Neuroscience}'', ``\textit{Nature Neuroscience}'', and ``\textit{Neuron}''). The selection of these journals relies on the fact that the resulting variability of their impact factors permits us to estimate its effect on citations. The first data set contained all articles ($n_E$ = 4,646) that were downloaded from Sci-hub between September 2015 and February 2016 \cite{Bohannon2016b, Bohannon2016c}. The second data set was extracted from the Scopus database and contained all articles published in the selected journals within the same time period. From this data set, we excluded articles already present in the first data set. For each journal, we next selected an equivalent number of articles as were included in the first data set. After the selection, the second data set had a total of 4,015 articles which we used in the analysis. Since this data set included only articles that had not been downloaded from Sci-hub, we use it as a control group. Thus, our design include an experimental and a control group framed in a  quasi-experiment design \cite{Antonakis2010} with a sample size of $N$ = 8,661. From full texts of the articles, we extracted the number of figures and tables in each paper, and we additionally obtained data regarding the H-index for both the junior and senior scientist of each paper, the impact factor of each journal, and the resources for conducting research in each author's country of affiliation (the method of data extraction is described in Section 1 of the appendix).

\subsection*{Statistical analyses}
Before testing the central hypothesis, we analyzed the statistical distribution of citations. Published articles with more than 2,000 citations proved to be outlier cases (see Figures S1 and S4 in the Appendix). To assess the robustness of results to different specifications of our model of interest (see equation \ref{eq1}), we employ a series of statistical modeling techniques inspired by the concept of a multiverse analysis \cite{steegen2016}. We therefore perform and report several modeling and statistical analyses suitable to the data at hand to assess systematic patterns in the results. To reach robust parameter estimates, we employed a series of analyses documented in sections 2A and 2B of the appendix. In particular, we assessed the sensitivity of relationships with regression diagnostics through Ordinary Least Squares, outliers influence, robust regression with instrumental variables, regressions with robust standard errors, and with heteroscedasticity errors, as well as with a series of generalized additive models for location, scale and shape \cite{Stasinopoulos2017}. The combination of these techniques allows us to control relevant confounding factors biasing the prediction of article citations (e.g., the length of a paper quantified by the number of pages and its possible relationship with the number of tables and figures, or the article type which was kept as a constant as all sampled papers were empirical). After controlling for problems of endogeneity, heteroscedasticity, and inconsistency, the interpretation of our robust estimates goes beyond claiming mere correlations or associations \cite{Antonakis2010}.

\section*{Results}
Papers downloaded from Sci-hub were cited 2.21 more frequently than articles not downloaded from Sci-hub (see table S2 in the appendix). Given the presence of outliers that could bias this estimate, we proposed a second model that was less sensitive to the influence of outliers. In the second model, the articles downloaded from Sci-hub had a citation rate of 1.16 higher than the papers without Sci-hub downloads (see table S4 in the appendix). As we detected that errors of this second model were not homoscedastic, we proposed a third model that corrected this source of bias. The results of this third model proved to be similar to those obtained in the first model (see table S5 in the appendix). In a fourth model, we ruled out alternative interpretations of our results by  introducing corrections that tackle the problems of endogeneity  \cite{Antonakis2010}. Here, the results showed that the citations of articles downloaded from Sci-hub were 1.954 times higher than the citations of articles without Sci-hub downloads. In all these models, we observed that the majority of the independent variables were significant predictors of citations, except the title length, the chaperon degree, the total number of tables included in a paper, and the resources of the affiliated country, as captured by the GDP per capita and Nature index. 

\begin{figure}[h!]
\centering
\includegraphics[width=0.6\textwidth]{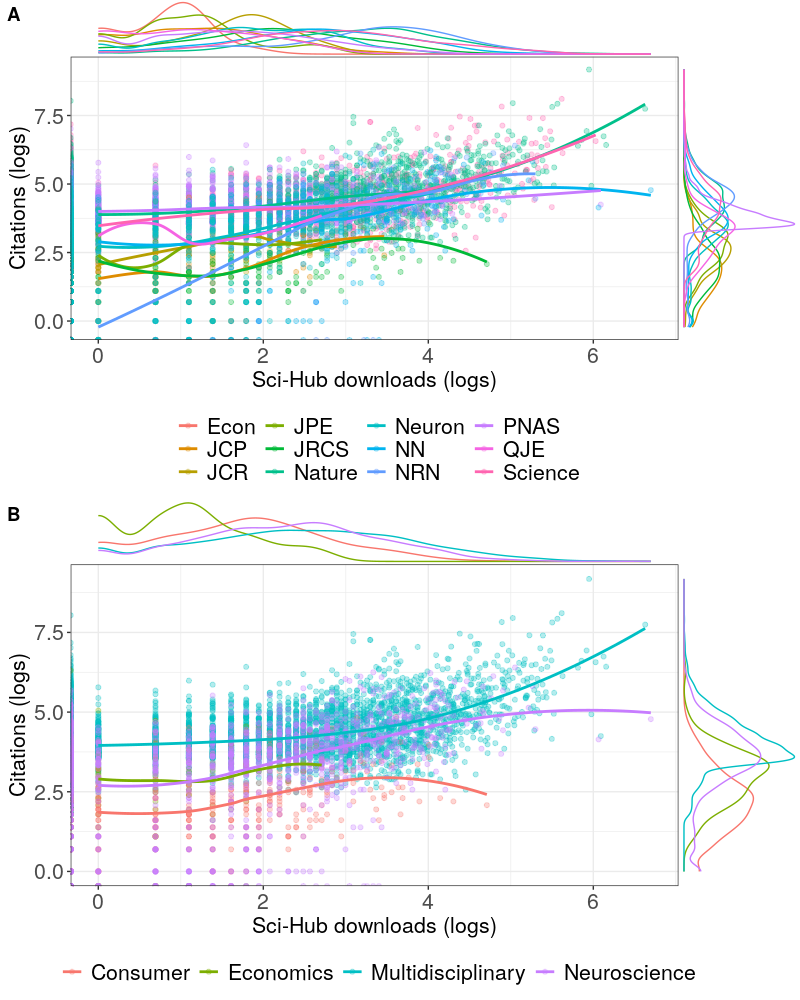}
\caption{The Sci-hub effect for (A) all analyzed journals (i.e., Econ refers to Econometrica; JPE is Journal of Political Economy; JCP is Journal of Consumer Psychology; JCR is Journal of Consumer Research; JRCS is Journal of Retailing and Consumer Services; NN is Nature Neuron; NRN is Nature Reviews Neuroscience; PNAS is Proceedings of the National Academy of Sciences of the United States of America; and QJE is Quarterly Journal of Economics) and (B) scientific disciplines.}
\label{fig1}
\end{figure}

To estimate a more fine-grained model, we proposed a fifth group of models. In these models we included a set of dummy variables as confounding factors, which allowed us to conduct a sensitivity analysis.
The estimation of the Sci-hub effect in this group of models varied between 1.19 and 1.412, and in each specification we observed that the total number of figures included in a paper, along with the number of authors per article, the H-index of the first author, and the impact factor of the journal, were significant predictors. A sixth group of models added a heteroscedasticity correction to the previous models (see table S8 in the appendix). Again, the estimation of the Sci-hub effect was significant, and regardless of the specification of the model, the effect fell between 2.215 and 2.460. A seventh group of models added corrections for the problems of endogeneity \cite{Antonakis2010}. After controlling all sources of bias, we found that papers downloaded from Sci-hub were cited 1.72 times more than papers not downloaded from Sci-hub (table S9 in the appendix). The Sci-hub effect was present in all selected journals and scientific disciplines that we analyzed (see Figure \ref{fig1}). 

\begin{figure}[h!]
\centering
\includegraphics[scale=.55]{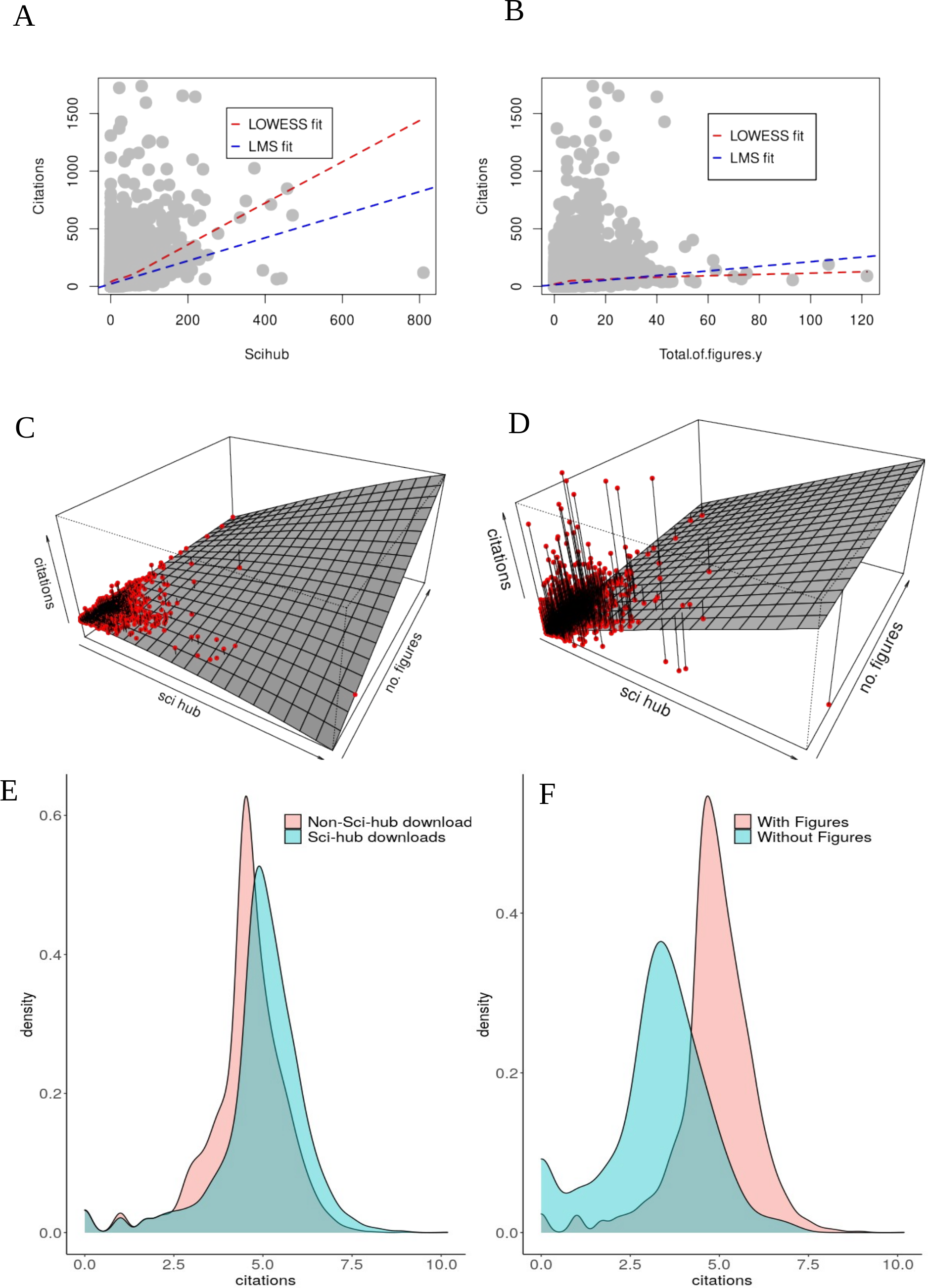}
\caption{Article citations as (A) a function of Sci-hub downloads, (B) a function of number of figures included in the article, (C-D) multivariate relationship between article citations and Sci-hub downloads with the number of figures included in a paper, (E) difference in the log-transformed statistical distributions of article citations as a function of Sci-hub downloads, and (F) difference in the log-transformed statistical distribution of article citations as a function of the presence/absence of figures in a published paper.}
\label{fig2}
\end{figure}

In all model specifications, we observed that the second best predictor of article citations, after the Sci-hub downloads, was the total number of figures in the paper (see Table S11 in the appendix). In the seventh group of models, we observed that papers with figures were 1.310 times more cited than those without figures (see Figure \ref{fig2}). 

The role of the impact factor of the journal where a paper was published varied with the exact specification of the model. Using a series of forward and backward stepwise variable selection procedure with a Generalized Additive Model for Location, Scale and Shape \cite{Stasinopoulos2017}, the impact factor of the journal ranked as the third strongest predictor of article citations (see table S11 in the appendix). The impact factor of a journal, however, was not consistently associated with the number of citations, as it was not statistically significant in different statistical model specifications (see tables S8 and S9 of the appendix).

\section*{Discussion}
The scientific impact of an article is often quantified by the number of citations received. Adding to the previous findings on this matter \cite{Radicchi2008,Bjork2012,Sinatra2016}, here we showed that articles accessed through Sci-hub systematically received more citations than papers not downloaded from Sci-hub. Up to the best of our knowledge, this is the first endeavor that shows an empirical estimate of the relationship between Sci-hub downloads and citations. In particular, this article provided a robust estimation of the effect of Sci-hub downloads on citations through a quasi-experimental design complemented with statistical methods that introduce corrections to control for other confounding factors \cite{Antonakis2010}. The Sci-hub effect has multiple implications. The first relates to the role of Sci-hub. Apart from being a global open-access research library \cite{Bohannon2016b}, we revealed its potential role as a data source that provides early indicators of the future impact of research. The number of downloads from Sci-hub could be potentially used for practical purposes instead of other indicators of quality and impact. The impact factor of the journal where an article is published is currently often used for this purpose. However, it is known that the impact factor of the journal suffers from low predictive validity and comes from a skewed distribution of citations, which is not highly informative about future citations of an individual article \cite{adler2009citation,hegarty2012consequences}. These limitations of the predictive value of the impact factor of a journal have been also supported by our study, given that the journal impact factor did not reliably predict future citations. 

The second implication of the Sci-hub effect revives an old concern regarding the scientific access and the political constraints to knowledge \cite{Horowitz1986}. While the lack of access to scientific articles is likely to hinder the work of researchers with limited access, the increasing use of Sci-hub \cite{Bohannon2016b} is related to the future dynamics of scientific knowledge dissemination \cite{Sa2019}. Any attempt of predicting how scientific knowledge will disseminate should include the response of universities, research funding agencies, and commercial publishers \cite{Oloughling2020}. The reaction of these institutions will affect researchers' decisions where to publish their work. While the option of divulging scientific works in open access journals is increasing in the last few decades \cite{Solomon2012,Solomon2014,Armstrong2015}, researchers are also becoming aware of the potential benefits of using scientific repositories (e.g., arXiv) for knowledge dissemination in the format of preprints \cite{Shuai2012,Berg2016}. We claim that Sci-hub, paradoxically, may help preserve the current publishing system because the lack of access to publications, which preprints and open access journals are trying to solve, may no longer be felt so strongly to find required increasing support.

Besides the Sci-hub effect, we also found that articles with a higher number of figures systematically receive more citations. This finding is particularly revealing from the role of graphs and tables as techniques to summarize results or depict relationships among variables in empirical articles. Given that the number of tables in a paper was not a significant predictor of its citations, the use of graphs in an article could not be driven by the ``hardness'' of a discipline or a similar factor that is shared between figures and tables \cite{Smith2002}. The effect must be driven by specific advantages of figures. As described in a seminal paper by Anscombe \cite{Anscombe1973}, graphs make understanding the information of a paper easier, and we show that the articles using them are more likely to be cited as a consequence. The decision to include graphs in an article might thus be one of the mechanisms that promote understanding and further use of scientific findings. Researchers could exploit this finding as a mechanism to maximize the impact of their research.

The dynamics of knowledge dissemination are tightly related to the access and use of scientific evidence. Our findings demonstrated two important factors that influence access. The Sci-hub effect revealed the importance of access by showing how Sci-hub almost doubles the citations of articles accessed through it. When the article itself is accessed, its findings also have to be properly conveyed, so that the publication can reach its full potential impact. Such an impact can be facilitated by summarizing the empirical information with graphs. These and other mechanisms that promote the use of scientific information is a topic that deserves further discussion in forthcoming studies.

\newpage
\section{Supplemental Material}
This appendix aims at providing detailed guidance for both understandings and reproducing the results of  ``\textit{The Sci-hub Effect: Sci-hub downloads lead to more article citation}.'' Section \ref{Sec2} describes the procedures we followed for data acquisition purposes. A first analysis is presented in Section \ref{Sec3}, which is composed of five subsections. In each of these subsections we provide the arguments that allow our readers understand how we reach more accurate estimates (i.e., robust estimates). Section \ref{Sec4} presents a second analysis with a similar structure and purpose as the preceding section. The difference between the first and the second analysis relies on the statistical techniques employed. The techniques employed in both analyses follow the recommendations from the perspective of multiverse analysis \cite{steegen2016}

\section{Data Acquisition}
\label{Sec2}
To estimate the empirical impact of Sci-Hub on the citations of papers, we used two data sets that we merged into a unified third data set. These data sets contain information about the articles published in a selected group of twelve journals. Three of these journals publish multidisciplinary research (i.e., ``\textit{Nature}'', ``\textit{Science}'', and ``\textit{Proceedings of the National Academy of Sciences}''). Other three are specialist journals in economics (i.e., ``\textit{The Quarterly Journal of Economics}'', ``\textit{Journal of Political Economy}'', and ``\textit{Econometrica}). We also sampled specialist journals in consumer research (i.e., ``\textit{Journal of Consumer Research}'', ``\textit{Journal of Retailing and Consumer Services}'', and ``\textit{Journal of Consumer Psychology}'') as well as specialist journals in neuroscience (``\textit{Nature Reviews Neuroscience}'', ``\textit{Nature Neuroscience}'', and ``\textit{Neuron}'').

The first data set, initially analyzed by Bohannon \cite{Bohannon2016b}, and shared by Bohannon and Elbakyan \cite{Bohannon2016c}, contains the information of the papers downloaded from Sci-hub between September 2015 and February 2016. This data set is available in the following repository (\url{https://datadryad.org/stash/dataset/doi:10.5061/dryad.q447c}), as well as in our own repository (\url{https://osf.io/7gvz2/}). The original data set contains a total of 27,819,966 rows out of which 10,373,380 corresponds to unique papers, published by 147,312 unique sources, including indexed journals, conference proceedings, etc. The second data set contains the information of all the articles published in the same journals and in the same period mentioned above. We obtained our second data set by conducting an advanced literature search inside Scopus database with the following search query string: ( ISSN ( 14764687 )  OR  ISSN ( 00280836 )  OR  ISSN ( 00368075 )  OR  ISSN ( 10916490 )  OR  ISSN ( 00278424 )  OR  ISSN ( 00935301 )  OR  ISSN ( 15375277 )  OR  ISSN ( 09696989 )  OR  ISSN ( 15327663 )  OR  ISSN ( 10577408 )  OR  ISSN ( 00335533 )  OR  ISSN ( 15314650 )  OR  ISSN ( 00129682 )  OR  ISSN ( 14680262 )  OR  ISSN ( 00223808 )  OR  ISSN ( 1537534x )  OR  ISSN ( 14710048 )  OR  ISSN ( 1471003x )  OR  ISSN ( 10976256 )  OR  ISSN ( 08966273 ) )  AND  ( LIMIT-TO ( PUBYEAR ,  2016 )  OR  LIMIT-TO ( PUBYEAR ,  2015 ) ). 

As a quality control, we performed a random sampling of all the articles retrieved, excluding those already present in the first data set. As in this second data set, the number of Sci-hub downloads is precisely equal to zero, we regard it as a control group ($n_C$ = 4,015) from which we are going to estimate comparisons for our experimental group ($n_E$ = 4,646). Thus, our design could be also regarded as a quasi-experiment \cite{Antonakis2010} with a reasonable sample size ($N$ = 8,661). With this sampling procedure, we merged both samples in one unified data set, and checked that articles' citation did not differ signficantly. The resulting second data set is available in our own repository (\url{https://osf.io/xb9yn/}). The third unified data set is also available in our own repository. After merging these data sets, we proceeded to download a full-text pdf version of all the articles using a python version of selenium software running in institutional access. Then, we counted the number of figures and tables, and obtained the H-index of the authors of each paper. To this purpose, we created a set of scripts in Bash, a Linux command language. In these scripts, we combined \textit{pdfgrep}, \textit{grep} and \textit{sed} commands to identify the occurrence of the words Figure and Table in the pdf files, discriminating supplementary material from the figures and tables published as a part of the manuscript. We customized these scripts to each journal.

Then, we used another bash script and pybliometrics, a python library, to automate the collection of data from the Scopus API. In particular, we used \textit{grep} to extract the H-Index and domain affiliation of all the authors for every single paper included in our final sample (n = 8,661). To estimate the Chaperone effect in scientific publishing \cite{Sekara2018}, we created a variable called ``Chaperone degree'' that mirrors the number of previous papers that the chaperone had published in the same journal, using another bash script. For example, if the last author (i.e., the chaperone) published two papers in \textit{Science} before the publication where they served as chaperone, then their chaperone degree equals two.

Because scientific productivity is systematically associated with the amount of resources in a country \cite{Laverde2019}, we used the domain affiliation of the authors to include two additional variables; namely, the annual GDP per capita and the Nature Index. Both of these variables are included for the first author (i.e, the junior scientist) and the last author (i.e., the chaperone) of each paper. The annual GDP per capita was obtained from World Bank, and the Nature Index is an author affiliation ranking based on the number of research articles published in a selected group of 82 high quality scientific journals. The Nature Index provides a close to real-time proxy of high-quality research output and collaboration at the institutional, national and regional level. Both of these metrics are of importance for controlling its potential effects as confounding variables in the relationships we want to explore. 

\section{Methods}
This section contains two parts. In both of these parts, we used a series of different statistical techniques to reach less biased estimates of relationships and discard all other confounding factors that could lead to a misleading interpretation of the results.

\subsection{PART 1}
\label{Sec3}
\begin{figure}[h!]
\centering
\includegraphics[width=0.45\textwidth]{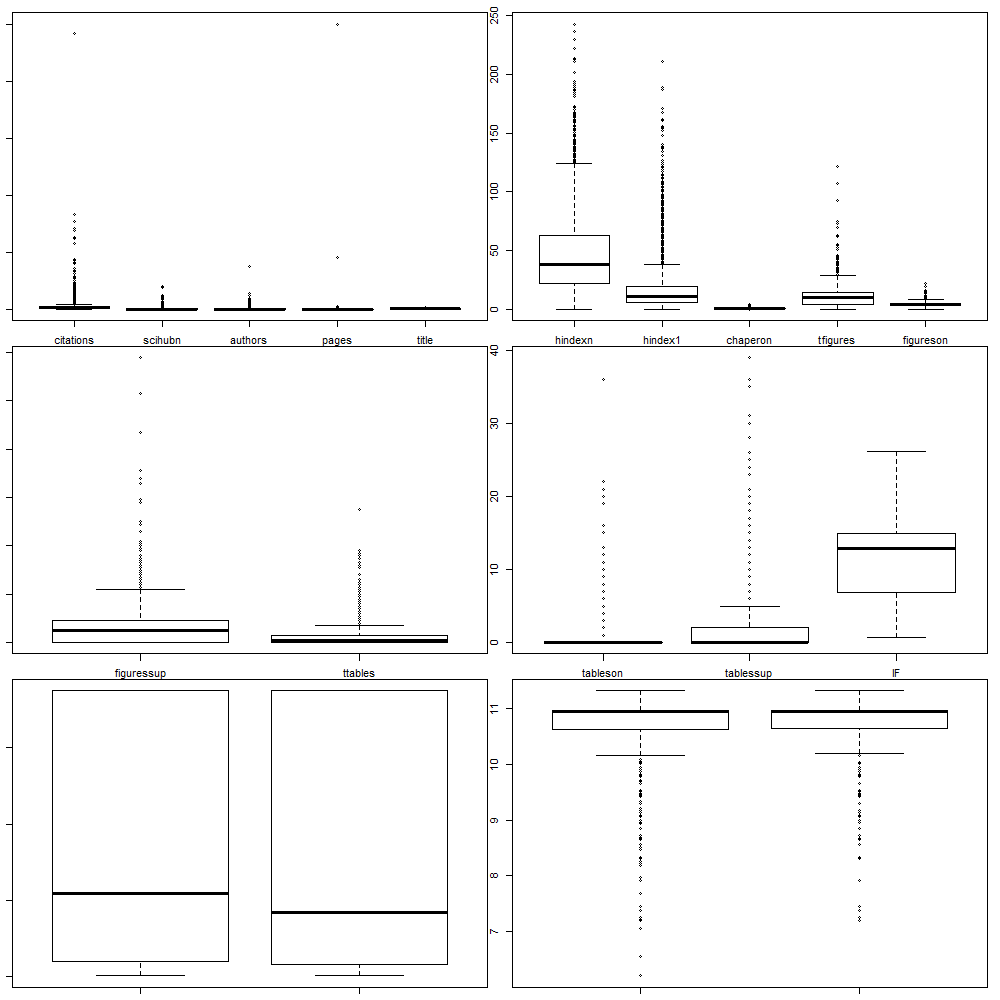}
\caption{Boxplot diagnostics}
\label{SI1}
\end{figure}

We cleaned the data set by omitting missing information. Then, the amount of data is reduced to 8,131 observations. We used these remaining observations to examine the behavior of outliers with box-plots and descriptive statistics. Both Figure \ref{SI1} and Table \ref{table1} reveal the presence of extreme values, particularly for citations, the number of pages, and the number of figures and tables. The presence of these outliers leads us to evaluate their influence on the regression models that are presented in the following subsections.

\begin{table}[h]
\centering
\caption{Descriptive Statistics}
\label{table1}
\resizebox{14cm}{!} {
\begin{tabular}{lcccccccccc}
\hline
                         & Citations              & Sci-hub             & Authors              & Pages                 & Title             & H-indexn                    & H-index1                   & Chaperon             & Tfigures             & Figureson            \\ \cline{2-11} 
Minimum                  & 0.00                   & 0.00                & 1.00                 & 1.00                  & 1.00              & 0.00                        & 0.00                       & 0.00                 & 0.00                 & 0.00                 \\
1st Quartile             & 29.00                  & 0.00                & 3.00                 & 4.00                  & 8.00              & 22.00                       & 6.00                       & 1.00                 & 4.00                 & 3.00                 \\
Median                   & 46.00                  & 1.00                & 6.00                 & 6.00                  & 11.00             & 38.00                       & 11.00                      & 1.00                 & 10.00                & 4.00                 \\
Mean                     & 79.28                  & 12.25               & 8.88                 & 8.00                  & 10.98             & 46.61                       & 16.67                      & 0.98                 & 9.41                 & 3.92                 \\
3rd Quartile             & 85.00                  & 12.00               & 10.00                & 8.00                  & 13.00             & 63.00                       & 19.00                      & 1.00                 & 14.00                & 5.00                 \\
Maximun                  & 9679.00                & 810.00              & 1484.00              & 10005.00              & 67.00             & 243.00                      & 211.00                     & 4.00                 & 122.00               & 22.00                \\ \cline{2-11} 
                         & Figuressup             & TTables             & Tableson             & Tablessup             & IF                & NatureIndex\_A1             & NatureIndex\_A2            & GDPpc\_A2            & GDPpc\_A2            &                      \\ \cline{2-11} 
Minimum                  & 0.00                   & 0.00                & 0.00                 & 0.00                  & 0.69              & 1.00                        & 1.00                       & 6.21                 & 7.20                 &                      \\
1st Quartile             & 0.00                   & 0.00                & 0.00                 & 0.00                  & 6.81              & 38.00                       & 30.00                      & 10.63                & 10.65                &                      \\
Median                   & 5.00                   & 1.00                & 0.00                 & 0.00                  & 12.87             & 218.00                      & 167.00                     & 10.95                & 10.95                &                      \\
Mean                     & 5.49                   & 1.99                & 0.51                 & 1488.00               & 11.97             & 354.00                      & 331.00                     & 10.69                & 10.75                &                      \\
3rd Quartile             & 9.00                   & 3.00                & 0.00                 & 2.00                  & 14.87             & 750.00                      & 750.00                     & 10.97                & 10.97                &                      \\
Maximun                  & 118.00                 & 55.00               & 36.00                & 39.00                 & 26.14             & 750.00                      & 750.00                     & 11.32                & 11.32                &                      \\ \hline
\multicolumn{11}{l}{\begin{tabular}[c]{@{}l@{}}Note: Citations = number of citations; Sci-hub; Authors = number of authors per document; Pages = number of pages; Title; H-indexn; H-index1; \\ Chaperon; Tfigures; Figureson; Figuressup; TTables; Tableson; Tablessup; IF; NatureIndex\_A1; NatureIndex\_A2; GDPpc\_A2; GDPpc\_A2\end{tabular}}
\end{tabular}}
\end{table}

\subsubsection{Regression diagnostics}

\begin{figure}[h!]
\centering
\includegraphics[width=0.37\textwidth]{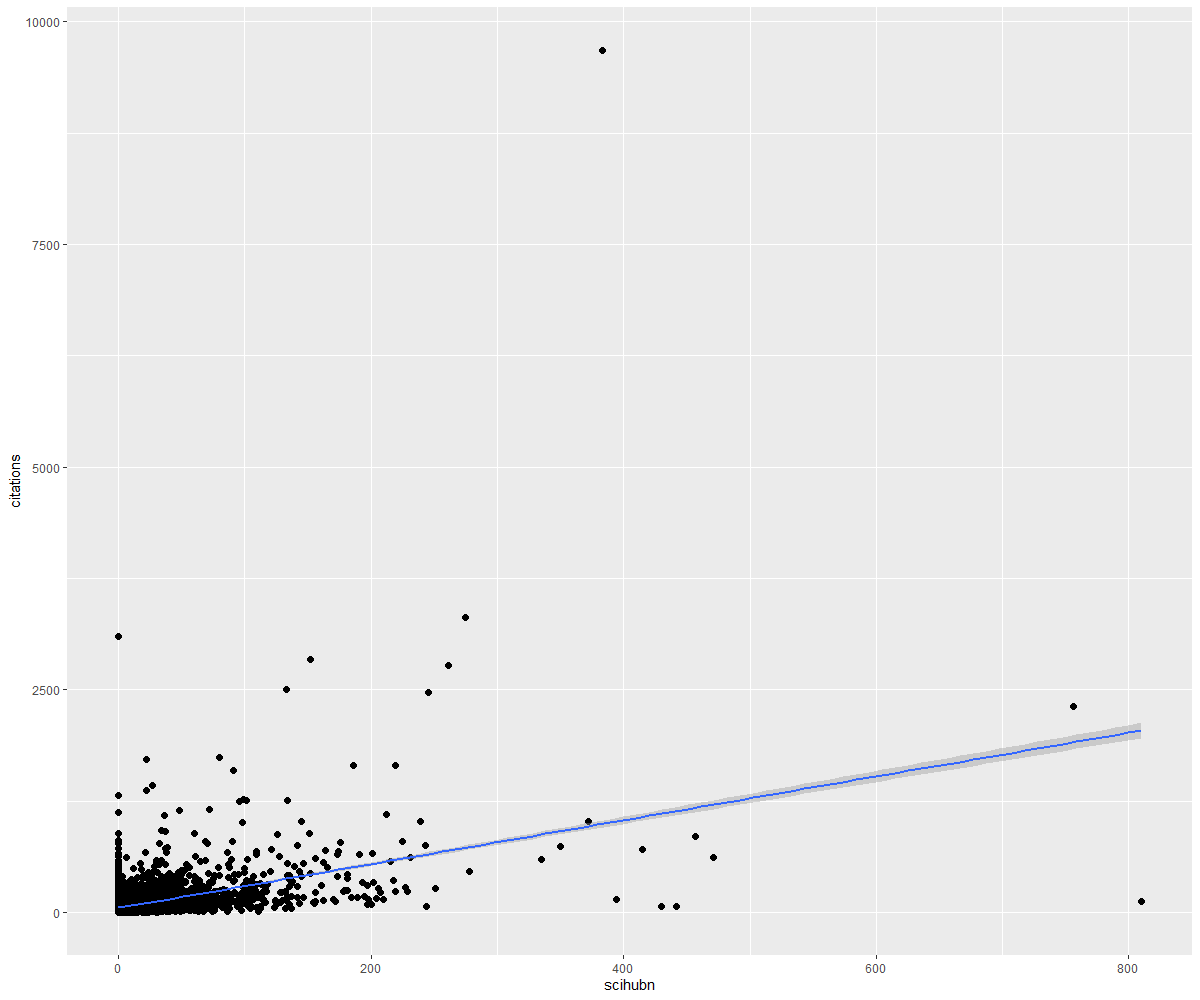}
\caption{Scatterplot diagnostic}
\label{SI2}
\end{figure}

An additional scatterplot analysis reveals a positive relationship between the number of citations and the number of Sci-hub downloads  (See figure, \ref{SI2}). This relationship, however, can be distorted by the outliers already mentioned.

Table \ref{table2} summarizes the results of a multiple regression. There is a positive and significant relationship between the number of Sci-hub downloads and the number of citations. Number of citations is also positively associated with the number of figures, authors per article, impact factor, and the H-index of both the first and last author of each paper. The length of the title has a negative relationship with the number of citations (i.e., papers with lengthy titles tend to have fewer citations), but this impact proved to be non-significant. The chaperone effect, the number of pages, the total number of tables, and the country resources, as captured by the GDP per capita and nature index, did not significantly predict the citations of a paper (at a statistical significance level of 5\%).  

\begin{table}[h!]
\centering
\caption{OLS Model}
\label{table2}
\begin{tabular}{lcccc}
\hline
& Estimate & Std. Error & t      & p-value: \\ \cline{2-5} 
Sci-hub                 & 2.221    & 0.055      & 40.405 & 0.000    \\
Authors Per Article     & 1.055    & 0.073      & 14.409 & 0.000    \\
Pages                   & -0.002   & 0.015      & -0.116 & 0.908    \\
Title Length            & -0.590   & 0.412      & -1.432 & 0.152    \\
H-indexn                & 0.127    & 0.059      & 2.142  & 0.032    \\
H-index1                & 0.736    & 0.101      & 7.309  & 0.000    \\
Chaperon Degree         & 5.504    & 3.261      & 1.688  & 0.092    \\
Total Figures           & 1.576    & 0.263      & 5.992  & 0.000    \\
Total Tables            & 0.143    & 0.522      & 0.275  & 0.784    \\
Impact Factor (IF)      & 1.804    & 0.343      & 5.254  & 0.000    \\
GDP per capita Author 1 & -13.430  & 35.500     & -0.378 & 0.705    \\
GDP per capita Author 2 & -2.027   & 4.366      & -0.464 & 0.642    \\
Nature Index Author 1   & 0.009    & 0.006      & 1.624  & 0.104    \\
Nature Index Author 2   & -0.007   & 0.006      & -1.247 & 0.213    \\
Intercept               & 23.322
   & 42.575     & 0.548  & 0.584    \\
$R^2$                      & 0.241    &            &        &          \\
Adjusted $R^2$             & 0.239    &            &        &          \\
F-statistic             & 183.500  &            &        &          \\
p-value:                & 0.000    &            &        &          \\ \hline
\end{tabular}
\end{table}

\begin{figure}[h!]
\centering
\includegraphics[width=0.35\textwidth]{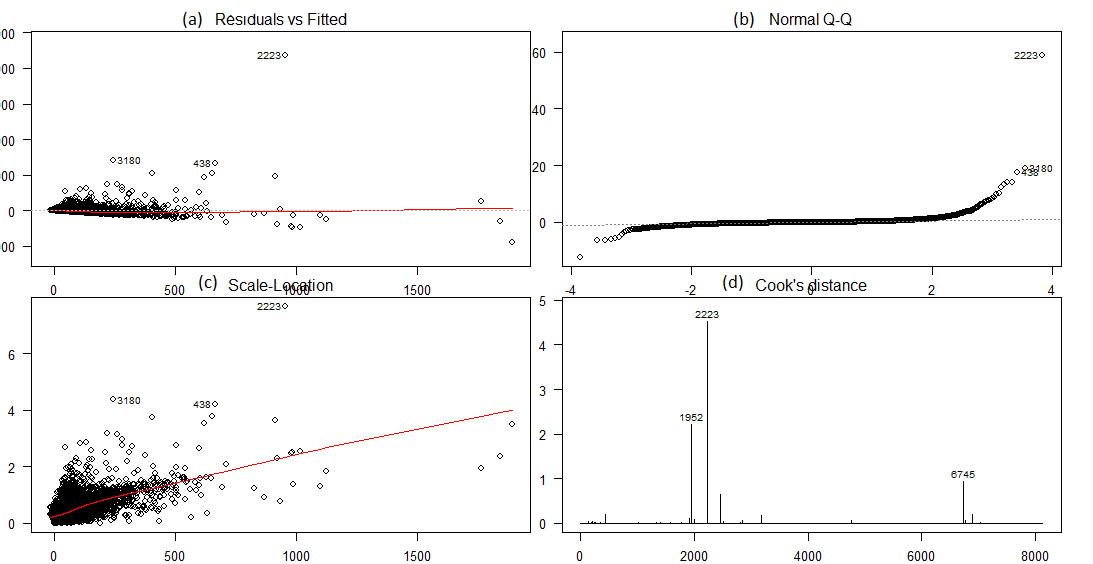}
\caption{Plot diagnostics: (a) Residuals vs Fitted values; (b) Normal Q-Q; (c) Scale-Location; (d) Cook's distance}
\label{fig3}
\end{figure}

Based on the results of Table \ref{table2}, we can make some regression diagnostics (e.g., the fulfillment of the assumptions of the model and the effects of the outliers on the results). These diagnostics allow us to decide the type of parameter estimation method that best suits the data. We begin by conducting a residual analysis to test the following assumption: $E(\epsilon | X)=0$. To do this, we depict the residuals against fitted values in Panel (a) of Figure \ref{fig3}. Individual estimates in this graph must be interpreted by comparing their distance to zero (i.e., the larger the distance from zero, the worse its estimate). We observe that some values can significantly alter the results of regressions. Our second analysis focuses on testing the normality of the residuals through a Q-Q plot, as depicted in Panel (b) of Figure \ref{fig3}. We notice that in both tails, several points do not fit the line, invalidating the results of the regressions (in particular, the confidence intervals and the significance tests). In Panel (c) in Figure \ref{fig3} we evaluate the i.i.d. assumption, particularly that of homoscedasticity. We notice that points are over the red line, indicating that the residuals have uniform variance. Again, the outlier points undermine this relationship, implying problems of heteroscedasticity. Finally, Cook's distance shows us that some points are very far from their average, as captured by Panel (d) in Figure \ref{fig3}.

In Table \ref{table3}, we used a deletion diagnostic to identify which influential observations may cause a substantial change in the fit when they are excluded from the model. We used the following measures of influence when $ith$ observation is deleted: a) DFFIT (how much the regression function changes), b) DFBETA (how much the coefficients change), c) COVRATIO (how much the covariance matrix change), d) $D^2$ (Cook’s distance, how much the entire regression function changes), and e) hat-values (for detecting high-leverage observations). In the literature, it is common to point out that an observation is considered unusual if it is detected by at least one of the aforementioned influence measures. Although many observations meet this condition, we only show a few for the sake of brevity. We noticed that observations such as 1,952 or 2,223 stand out using any measure of influence, demonstrating how detrimental these points can be for the results of the regression analysis. 

\begin{table}[h!]
\scriptsize
\centering
\caption{Analysis of the influence of outliers}
\label{table3}
\resizebox{14cm}{!} {
\begin{tabular}{cccccccccc}
\hline
Obs  & dfb.schb & dfb.athr    & dfb.pags    & dfb.titl   & dfb.hn     & dfb.h1 & dfb.chpr & dfb.tfgrs & dfb.tbls \\ \hline
438  & 1.74*    & -0.15       & 0.01        & -0.08      & -0.08      & -0.11  & 0.02     & -0.12     & -0.12    \\
637  & 0.01     & 0           & 0           & 0          & 0.03       & -0.08  & -0.02    & -0.01     & 0        \\
1952 & 0.2      & -5.77\_*    & -0.02       & 0.18       & 0.17       & 0.27   & -0.05    & 0.58      & 0.26     \\
2223 & 10.24*   & -1.07*      & 0.08        & -0.92      & -0.47      & 1.35*  & 0.43     & -0.8      & -0.46    \\
6745 & -3.77*   & 0.24        & -0.02       & -0.06      & 0.21       & 0.12   & 0.07     & 0.15      & -0.3     \\ \hline
Obs  & dfb.IF   & dfb.GDP\_A1 & dfb.GDP\_A2 & dfb.NI\_A1 & dfb.NI\_A2 & dffit  & cov.r    & cook.d    & hat      \\ \hline
438  & -0.28    & 0.03        & -0.14       & 0.11       & 0.15       & 1.79*  & 0.56*    & 0.21      & 0.01*    \\
637  & 0.01     & 0           & 0           & 0          & 0.01       & -0.09  & 1.01*    & 0         & 0.01*    \\
1952 & 0.3      & -0.03       & -0.08       & 0.2        & 0.03       & -5.78* & 1.94*    & 2.22*     & 0.51*    \\
2223 & -1.09*   & 1.07*       & -0.76       & 1.70*      & -1.50*     & 10.86* & 0.00*    & 4.52*     & 0.02*    \\
6745 & 0.6      & -0.14       & -0.06       & 0.05       & 0          & -3.80* & 0.82*    & 0.95      & 0.08*    \\ \hline
\multicolumn{10}{l}{Notes: dfb = changes in coefficients; dffit = change in fitted values; cov.r = change in covariance matrix;}\\
\multicolumn{10}{l}{ cook.d = Cook’s distance reduce information to a single value per observation; hat = hatvalues.}
\end{tabular}
}
\end{table}

\subsubsection{Dealing with outliers}
So far, we found the presence of outliers that threaten the validity of traditional regression analyses. One possible solution, given the study of the regression diagnosis, would consist of eliminating the problematic observations. However, with this technique, valuable information is lost. Instead, we use a robust regression that is less sensitive to outliers \cite{rousseeuw2005robust}. Table \ref{table4} shows the results for the model presented in Table \ref{table2} estimated by robust regression, through the use of iterated re-weighted least squares.            

\begin{table}[h!]
\centering
\scriptsize
\caption{Robust Regression}
\label{table4}
\begin{tabular}{lcccc}
\hline
                        & Estimate & Std. Error & t      & p-value: \\ \cline{2-5} 
Sci-hub                 & 1.160    & 0.014      & 86.026 & 0.000    \\
Authors Per Article     & 0.853    & 0.018      & 47.534 & 0.000    \\
Pages                   & -0.002   & 0.004      & -0.484 & 0.628    \\
Title Length            & -0.062   & 0.101      & -0.609 & 0.543    \\
H-indexn                & 0.155    & 0.015      & 10.676 & 0.000    \\
H-index1                & 0.233    & 0.025      & 9.437  & 0.000    \\
Chaperon Degree         & 0.484    & 0.800      & 0.605  & 0.545    \\
Total Figures           & 1.349    & 0.065      & 20.921 & 0.000    \\
Total Tables            & 0.290    & 0.128      & 2.267  & 0.023    \\
Impact Factor (IF)      & 1.197    & 0.084      & 14.214 & 0.000    \\
GDP per capita Author 1 & -1.081  & 0.894      & -1.209 & 0.227    \\
GDP per capita Author 2 & 0.457    & 1.071      & 0.427  & 0.669    \\
Nature Index Author 1   & -0.001   & 0.001      & -0.572 & 0.567    \\
Nature Index Author 2   & -0.002   & 0.001      & -1.419 & 0.156    \\
Intercept               & 23.187   & 17.476     & 1.327  & 0.185    \\
F-statistic             & 617.85   &            &        &          \\
p-value:                & 0.0000   &            &        &          \\
Observations            & 8129     &            &        &          \\ \hline
\end{tabular}
\end{table}

Robust regression assigns a higher weighting to observations that generate a lower residual. Comparing the results of OLS and Robust regressions shows that coefficients, signs, and statistical significance are very different, revealing a strong influence of outliers on model parameters in the OLS regression.

\subsubsection{Dealing with heteroscedasticity}

An important assumption in traditional regression models is that errors must be homoscedastic. The violation of this assumption can lead to the use of covariance matrix estimators that can be inconsistently estimated. Although a first exploration was already carried out through graphical analysis, we test this assumption in our model through the Breusch-Pagan test. As expected in the cross-section models, the test shows the presence of heteroscedasticity problems ($BP = 283.76, df = 14, p-value < 2.2e-16$) whose solution consists of employing heteroscedasticity-consistent estimators, through the Huber-White basic sandwich estimator.  Table \ref{table5} shows the results for regression with robust standard errors.  

\begin{table}[h!]
\centering
\caption{Regression with Robust Standard Errors}
\label{table5}
\begin{tabular}{lcccc}
\hline
                        & Estimate & \begin{tabular}[c]{@{}c@{}}Robust \\ Std. Error\end{tabular} & t      & p-value: \\ \cline{2-5} 
Sci-hub                 & 2.221    & 0.496                                                        & 4.470  & 0.000    \\
Authors Per Article     & 1.055    & 0.352                                                        & 3.000  & 0.003    \\
Pages                   & -0.002   & 0.001                                                        & -1.390 & 0.164    \\
Title Length            & -0.590   & 0.366                                                        & -1.610 & 0.107    \\
H-indexn                & 0.127    & 0.066                                                        & 1.910  & 0.056    \\
H-index1                & 0.736    & 0.149                                                        & 4.920  & 0.000    \\
Chaperon Degree         & 5.499    & 2.807                                                        & 1.960  & 0.050    \\
Total Figures           & 1.575    & 0.319                                                        & 4.940  & 0.000    \\
Total Tables            & 0.143    & 0.478                                                        & 0.300  & 0.765    \\
Impact Factor (IF)      & 1.804    & 0.419                                                        & 4.310  & 0.000    \\
GDP per capita Author 1 & 0.009    & 0.008                                                        & 1.110  & 0.265    \\
GDP per capita Author 2 & -0.007   & 0.079                                                        & -0.930 & 0.351    \\
Nature Index Author 1   & -13.430  & 43.875                                                       & -0.310 & 0.760    \\
Nature Index Author 2   & -2.027   & 4.271                                                        & -0.470 & 0.635    \\
Intercept               & 23.322   & 30.391                                                       & 0.767  & 0.443    \\
F-statistic             & 53.490   &                                                              &        &          \\
p-value:                & 0.000    &                                                              &        &          \\
Observations            & 8131     &                                                              &        &          \\ \hline
\end{tabular}
\end{table}

With this correction, the results are similar to the results of OLS with regard to the sign, magnitude, and statistical significance of the coefficients, but different from those of a robust regression in the size of the coefficients.

\subsubsection{Dealing with endogeneity}

Another assumption in OLS models is that of endogeneity, which takes place when one of the independent variables is related to the residual term in the regression equation. In that case, the OLS estimates can be spurious. The traditional technique to correct this problem is using instrumental variables. However, the application of this method needs to generate external instruments that are not always available. Here, we rely on Lewbel's methodology to evaluate the endogeneity problem \cite{lewbel2012using}. Although our results are based on R packages, the application of the Lewbel's methodology is best developed in Stata, particularly the tests of overidentifying restrictions. Sargan’s statistic  is not robust  in the presence of conditional heteroskedasticity, so we rely on Hansen J statistic. Table \ref{table6} shows the results obtained through the Lewbel's method, assuming that Sci-hub and Nature Index variables are endogenous. The Hansen test allows us to test the orthogonality conditions for the instruments. The results mentioned above indicate that the model may have an endogeneity problem, so it needs to be instrumented in different models. It should be clear that the tests of assumptions of OLS models allow us to conclude that our proposed regression model is affected by outliers, heterosdasticity and endogeneity problems. To overcome these problems, our results will be presented using robust regression, regression with robust standard errors, and instrumented variables based on heteroscedasticity.   

\newpage

\begin{table}[h!]
\centering
\caption{Regression with instrumented variables based on Heterocedasticity}
\label{table6}
\begin{tabular}{lcccc}
\hline
                             & Estimate & Std. Error & t      & p-value: \\ \cline{2-5} 
Sci-hub                      & 1.954    & 0.182      & 10.740 & 0.000    \\
Authors Per Article          & 0.753    & 0.228      & 3.310  & 0.001    \\
Pages                        & -0.003   & 0.001      & -2.460 & 0.014    \\
Title Length                 & -0.212   & 0.207      & -1.030 & 0.305    \\
H-indexn                     & 0.171    & 0.046      & 3.750  & 0.000    \\
H-index1                     & 0.393    & 0.081      & 4.850  & 0.000    \\
Chaperon Degree              & 2.560    & 2.064      & 1.240  & 0.215    \\
Total Figures                & 1.598    & 0.192      & 8.320  & 0.000    \\
Total Tables                 & 0.191    & 0.315      & 0.600  & 0.545    \\
Impact Factor (IF)           & 1.794    & 0.219      & 8.180  & 0.000    \\
GDP per capita Author 1      & -12.378  & 35.055     & -0.350 & 0.724    \\
GDP per capita Author 2      & 0.151    & 4.046      & 0.040  & 0.970    \\
Nature Index Author 1        & 0.003    & 0.020      & 0.150  & 0.878    \\
Nature Index Author 2        & 0.001    & 0.022      & 0.050  & 0.959    \\
F-statistic                  & 62.89    &            &        &          \\
p-value:                     & 0.000    &            &        &          \\
Observations                 & 8131     &            &        &          \\
$R^2$                           & 0.2337   &            &        &          \\
Hansen J statistic (p-value) & 0.1666   &            &        &          \\ \hline
\end{tabular}
\end{table}

The results of Table \ref{table6} show that the majority of variables proved to be significant predictors of citations, except the title length, the chaperon degree, the total tables, and the resources of the affiliated country, as captured by the GDP per capita and Nature index.

\subsubsection{Results and Robustness analysis}

In this section, we present our final results and test the robustness of them by using different sets of models and methods. The following equation gives the specification we are trying to estimate:
\begin{equation}
C_i = \beta_i \times SciHub_i + X_i^{'} \gamma_i + \sum_{j=1}^{4}\delta_{ij}\times discipline_{ij} + \sum_{k=1}^{12}\varphi_{ik}\times journal_{ik} + \theta_i
\label{eq1}
\end{equation}

Where $C_i$ stands for the number of citations the paper $i$ has received, $\beta$ is our parameter of interest as it quantifies the relationship between the citation of a paper and the number of times the paper $i$ was downloaded through SciHub; $X'$ is a vector containing the following control variables: The impact factor of the journal where the paper was published; the length of the title of the paper, as captured by the number of types or unique words in it; the number of graphs included in the paper $i$ for communicating scientific findings, the number of tables included in the paper $i$; the chaperone effect captured by the H-index of the first and last author of paper $i$; the number of authors of the paper $i$. $\theta_i$ represents the residuals of our model. 
A reasonable assumption in our model would be that each discipline and journal have different citation patterns. Given the variability intrinsically associated with the scientific discipline and the particular journal where the paper was published, we also include dummies for $discipline$ and $journal$ type to control for hidden confounds. The above specification could be understood as an extended specification of equation \ref{eq1} in the main manuscript. Table \ref{table7} shows the results of the estimates of robust regression.      
\newpage
\begin{table}[h!]
\scriptsize
\centering
\caption{Effects of Sci-Hub on citations based on Robust Regression}
\label{table7}
\begin{tabular}{lcccccc}
\hline
                        & (1)       & (2)      & (3)      & (4)      & (5)       & (6)      \\ \cline{2-7} 
Sci-hub                 & 1.412***  & 1.302*** & 1.224*** & 1.190*** & 1.306***  & 1.150*** \\
                        & (0.014)   & (0.013)  & (0.013)  & (0.013)  & (0.013)   & (0.013)  \\
Figures                 &           &          & 1.319*** &          &           & 1.180*** \\
                        &           &          & (0.061)  &          &           & (0.063)  \\
Tables                  &           &          & 0.660*** &          &           & 0.465*** \\
                        &           &          & (0.130)  &          &           & (0.128)  \\
Pages                   &           &          & -0.000        &          &           & 0        \\
                        &           &          & (0.004)  &          &           & (0.003)  \\
Title Length            &           &          & 0.019    &          &           & -0.082   \\
                        &           &          & (0.097)  &          &           & (0.095)  \\
Authors Per Article     &           &          &          & 0.904*** &           & 0.750*** \\
                        &           &          &          & (0.017)  &           & (0.017)  \\
Hindexn                 &           &          &          & 0.137*** &           & 0.121*** \\
                        &           &          &          & (0.014)  &           & (0.014)  \\
H-index1                &           &          &          & 0.052**  &           & 0.166*** \\
                        &           &          &          & (0.024)  &           & (0.024)  \\
Chaperon Degree         &           &          &          & 1.481*   &           & -0.065   \\
                        &           &          &          & (0.777)  &           & (0.756)  \\
Impact Factor (IF)      &           &          &          &          & 1.468**   & 1.427**  \\
                        &           &          &          &          & (0.740)   & (0.695)  \\
GDP per capita Author 1 &           &          &          &          & -15.426*  & -1.599   \\
                        &           &          &          &          & (8.644)   & (8.142)  \\
GDP per capita Author 2 &           &          &          &          & 1.899*    & 1.22     \\
                        &           &          &          &          & (1.057)   & (1.001)  \\
Nature Index Author 1   &           &          &          &          & -0.002*   & -0.002   \\
                        &           &          &          &          & (0.001)   & (0.001)  \\
Nature Index Author 2   &           &          &          &          & -0.004*** & -0.002*  \\
                        &           &          &          &          & (0.001)   & (0.001)  \\
Intercept               & 43.053*** & -3.163   & -9.666   & -7.191   & -4.859    & -39.939* \\
                        & (0.483)   & (9.796)  & (9.409)  & (9.493)  & (22.324)  & (21.057) \\
Control dummies         & No        & Yes      & Yes      & Yes      & Yes       & Yes      \\
F-statistic             & 4968.15   & 587.81   & 1034.55  & 633.57   & 445.05    & 415.71   \\
p-value:                & 0         & 0        & 0        & 0        & 0         & 0        \\
Observations            & 8128      & 8129     & 8129     & 8129     & 8130      & 8129     \\ \hline
\multicolumn{7}{l}{Notes: Estimation by mean of iterated re-weighted least squares (IRLS). ***$p<0.001$, **$p<0.05$,}\\
\multicolumn{7}{l}{*$p<0.1$. Standard errors in brackets.}
\end{tabular}
\end{table}

We run equation \ref{eq1} again. However, this time we introduce blocks of variables gradually to conduct a sensitivity analysis. First, model 1 does not include any control variables. Here, the number of times the paper was downloaded from Sci-hub has a positive and significant effect on the number of citations. The following model introduces the dummies by the type of discipline (i.e., multidisciplinary, economics, consumer, or neuroscience) and journal. In this case, the results for Sci-hub remain almost unchanged. In the third model, we added a series of variables related to the characteristics of the document (i.e., number of figures, tables, pages and the extension of the title). Once again, Sci-hub remains robust to this new specification. 

The number of figures and tables included in a paper both show a positive and significant relationship with the number of citations. Conversely, the pages and the length of the title show the opposite association. Next, we introduced a new block of variables related to the characteristics of the authors (i.e., the H-index of both the first and last author, the number of authors of the paper, and the chaperone degree). The introduction of these new variables does not change the results for the Sci-hub effect on article citations. All variables reveal positive and significant effects for citations except the chaperon degree (at a statistical significance level of 5\%). In model 5, we introduced variables related to the context in which the authors and journals operate (such as the GDP per capita for the country of the authors, the impact factor of the journal, and the nature index). In this model, the Sci-hub coefficient is still positive and highly significant. For the rest of the variables, only the impact factor seems to correspond to the expectations in terms of sign and statistical significance. Finally, we added all the control variables in the same model. The results remain unchanged from the previous specifications. In table \ref{table8}, we run the same models but now we estimate the parameters with a heteroscedasticity correction by using robust errors. 
\newpage

\begin{table}[h!]
\scriptsize
\centering
\caption{Effects of Sci-Hub on citations based on Robust Standard Errors}
\label{table8}
\begin{tabular}{lcccccc}
\hline
                        & (1)       & (2)       & (3)      & (4)        & (5)       & (6)      \\ \cline{2-7} 
Sci-hub                 & 2.460***  & 2.355***  & 2.321*** & 2.247***   & 2.354***  & 2.215*** \\
                        & (0.484)   & (0.498)   & (0.504)  & (0.501)    & (0.498)   & (0.504)  \\
Figures                 &           &           & 1.227*** &            &           & 1.414*** \\
                        &           &           & (0.376)  &            &           & (0.329)  \\
Tables                  &           &           & 0.783    &            &           & 0.123    \\
                        &           &           & (0.509)  &            &           & (0.525)  \\
Pages                   &           &           & -0.001   &            &           & -0.000   \\
                        &           &           & (0.002)  &            &           & (0.002)  \\
Title Length            &           &           & -0.549   &            &           & -0.664*  \\
                        &           &           & (0.421)  &            &           & (0.402)  \\
Authors Per Article     &           &           &          & 1.067***   &           & 1.035*** \\
                        &           &           &          & (0.356)    &           & (0.348)  \\
Hindexn                 &           &           &          & 0.127*     &           & 0.110    \\
                        &           &           &          & (0.068)    &           & (0.068)  \\
H-index1                &           &           &          & 0.587***   &           & 0.716*** \\
                        &           &           &          & (0.173)    &           & (0.164)  \\
Chaperon Degree         &           &           &          & 6.430**    &           & 4.996*   \\
                        &           &           &          & (2.718)    &           & (2.805)  \\
Impact Factor (IF)      &           &           &          &            & 3.585     & 3.126    \\
                        &           &           &          &            & (3.238)   & (3.237)  \\
GDP per capita Author 1 &           &           &          &            & -14.177   & 0.921    \\
                        &           &           &          &            & (45.648)  & (43.358) \\
GDP per capita Author 2 &           &           &          &            & 0.519     & -1.143   \\
                        &           &           &          &            & (4.306)   & (4.285)  \\
Nature Index Author 1   &           &           &          &            & 0.007     & 0.009    \\
                        &           &           &          &            & (0.008)   & (0.008)  \\
Nature Index Author 2   &           &           &          &            & -0.006    & -0.008   \\
                        &           &           &          &            & (0.008)   & (0.008)  \\
Intercept               & 49.156*** & -5.720*** & -7.528   & -17.707*** & -26.230   & -50.348  \\
                        & (4.772)   & (1.459)   & (4.705 ) & (3.676)    & (103.852) & (95.821) \\
Control dummies         & No        & Yes       & Yes      & Yes        & Yes       & Yes      \\
F-statistic             & 25.84     & 305.94    & 138.26   & 154.34     & 255.42    & 97.03    \\
p-value:                & 0.000     & 0.000     & 0.000    & 0.000      & 0.000     & 0.000    \\
Observations            & 8131      & 8131      & 8131     & 8131       & 8131      & 8131     \\ \hline
\multicolumn{7}{l}{Notes: ***$p<0.001$, $p<0.05$, *$p<0.1$. Standard errors Huber-White in brackets.}\\
\end{tabular}
\end{table}

Regardless of the specification we use, the results show that the effect of Sci-hub on the number of citations remains positive and significant. Concerning the characteristics of the document or the authors, the results vary for some variables. For example, variables such as the number of tables, the extension of the title, or the chaperon effect do not prove to be very robust to different specifications.

Finally, in Table \ref{table9}, we estimate our models by tackling the endogeneity problem. In general terms, the models show good performance. We were able to verify the validity of the instruments, except for models 1 and 3, when they were evaluated through the Hansen J statistic. As can be seen, the effect of Sci-hub on article citations remains robust to different specifications, while the other variables have a similar behavior to that of Table \ref{table8}, except for the variables related to the context in which the authors and journals operate (i.e., impact factor, author $i$'s GDP per capita, author $i$'s nature index). 
\begin{table}[h!]
\scriptsize
\centering
\caption{Effects of Sci-Hub on citations using Heteroskedasticity errors}
\label{table9}
\begin{tabular}{lccccc}
\hline
                             & (1)       & (2)       & (3)        & (4)      & (5)      \\ \cline{2-6} 
Sci-hub                      & 1.893***  & 2.012***  & 1.755***   & 1.940*** & 1.721*** \\
                             & (0.190)   & (0.169)   & (.1611558) & (0.138)  & (0.137)  \\
Figures                      &           & 1.447***  &            &          & 1.310*** \\
                             &           & (0.200)   &            &          & (0.196)  \\
Tables                       &           & 0.780**   &            &          & 0.370    \\
                             &           & (0.354)   &            &          & (0.316)  \\
Pages                        &           & -0.001    &            &          & -0.001   \\
                             &           & (0.002)   &            &          & (0.002)  \\
Title Length                 &           & -0.174    &            &          & -0.205   \\
                             &           & (0.192)   &            &          & (0.207)  \\
Authors Per Article          &           &           & 0.948***   &          & 0.754*** \\
                             &           &           & (0.305)    &          & (0.211)  \\
Hindexn                      &           &           & 0.183***   &          & 0.142*** \\
                             &           &           & (0.056)    &          & (0.044)  \\
H-index1                     &           &           & 0.314***   &          & 0.355*** \\
                             &           &           & (0.087)    &          & (0.081)  \\
Chaperon Degree              &           &           & 2.989      &          & 2.449    \\
                             &           &           & (1.919)    &          & (1.994)  \\
Impact Factor (IF)           &           &           &            & 3.331**  & 2.276    \\
                             &           &           &            & (1.510)  & (1.816)  \\
GDP per capita Author 1      &           &           &            & -30.500  & -7.031   \\
                             &           &           &            & (25.334) & (28.881) \\
GDP per capita Author 2      &           &           &            & 2.454    & 0.860    \\
                             &           &           &            & ( 2.650) & (3.042)  \\
Nature Index Author 1        &           &           &            & -0.004   & 0.001    \\
                             &           &           &            & (0.006)  & (0.009)  \\
Nature Index Author 2        &           &           &            & -0.007   & 0.002    \\
                             &           &           &            & (0.006)  & (0.009)  \\
Intercept                    & -4.966*** & -7.746*** & -13.147*** & -0.289   & -51.176  \\
                             & (0.991)   & (2.327)   & (2.419)    & (56.040) & (67.757) \\
Control dummies              & Yes       & Yes       & Yes        & Yes      & Yes      \\
F-statistic                  & 315.77    & 190.35    & 182.54     & 220.61   & 65.91    \\
p-value:                     & 0.000     & 0.000     & 0.000      & 0.000    & 0.000    \\
$R^2$                           & 0.211     & 0.217     & 0.235      & 0.212    & 0.234    \\
Hansen J statistic (p-value) & 0.023     & 0.101     & 0.044      & 0.294    & 0.484    \\
Observations                 & 8131      & 8131      & 8131       & 8131     & 8131     \\ \hline
\multicolumn{6}{l}{Notes: ***$p<0.001$, $p<0.05$, *$p<0.1$. Robust standard errors in brackets. Models through Two-step GMM}
\end{tabular}
\end{table}

\subsection{PART 2}
\label{Sec4}
\begin{figure}[h!]
\centering
\includegraphics[width=0.4\textwidth]{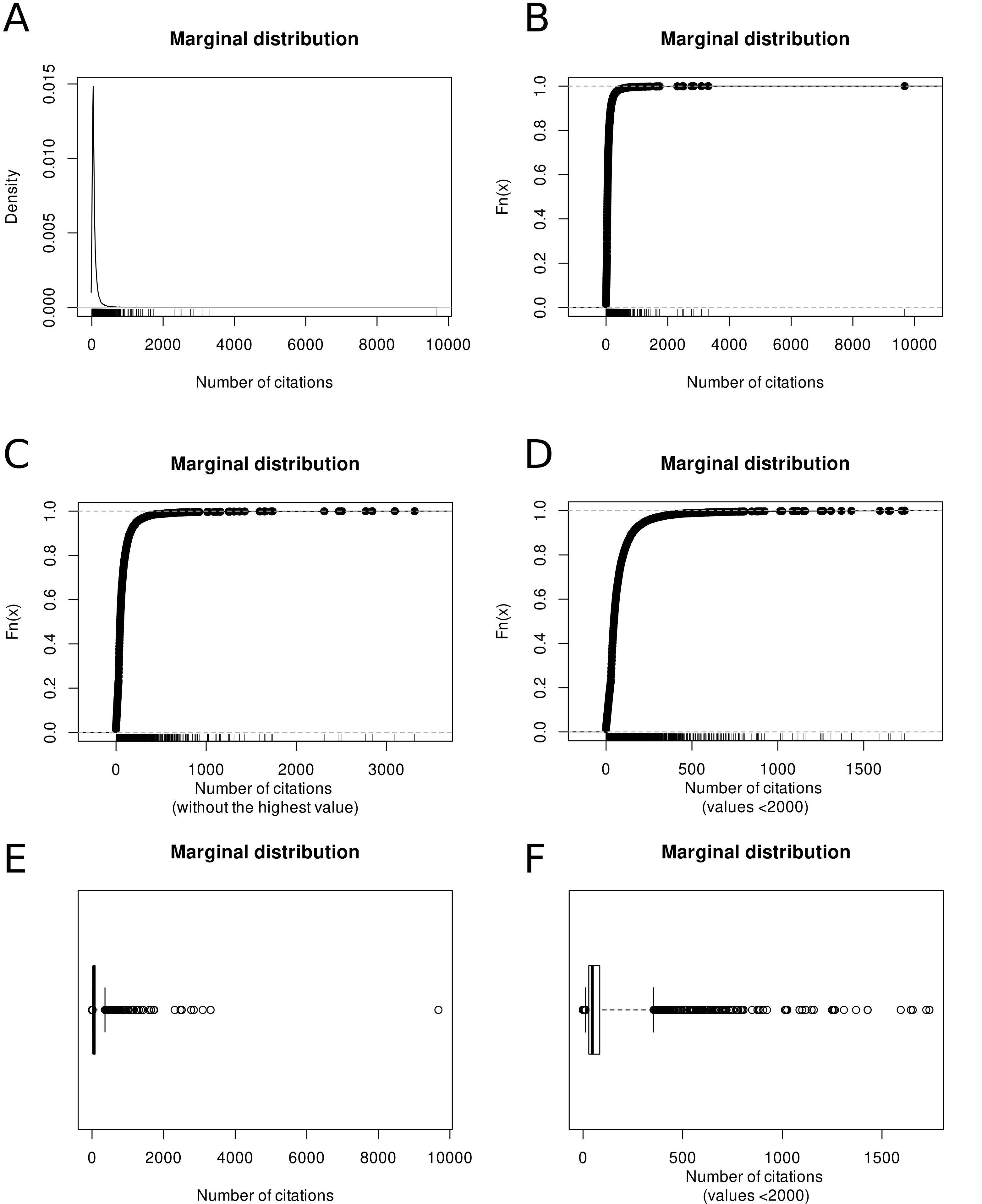}
\caption{Density (A) and empirical cumulative distribution function of the original data set (B), and after excluding citation values higher than 9000 (C) and higher than 2000 (D). The last row displays adjusted boxplots of the orginal dataset (E) and after excluding articles with citations $>$2000 (F).}
\label{fig4}
\end{figure}

This second analysis offers a different approach as compared to those described in the preceding section. We begin the analysis by focusing on the marginal distribution of data. Figure \ref{fig4}A depicts the complete distribution. The presence of an outlier article with more than 9,000 citations is evident in Figure \ref{fig4}B. After excluding the article, it is clear that there are still some articles with more than 2,000 citations (see Figure \ref{fig4}C). By removing those articles, it is possible to obtain a smoother distribution (see Figure \ref{fig4}E and F).

\subsubsection{Fitting the shape of the marginal distribution}
\begin{figure}[h!]
\centering
\includegraphics[width=0.8\textwidth]{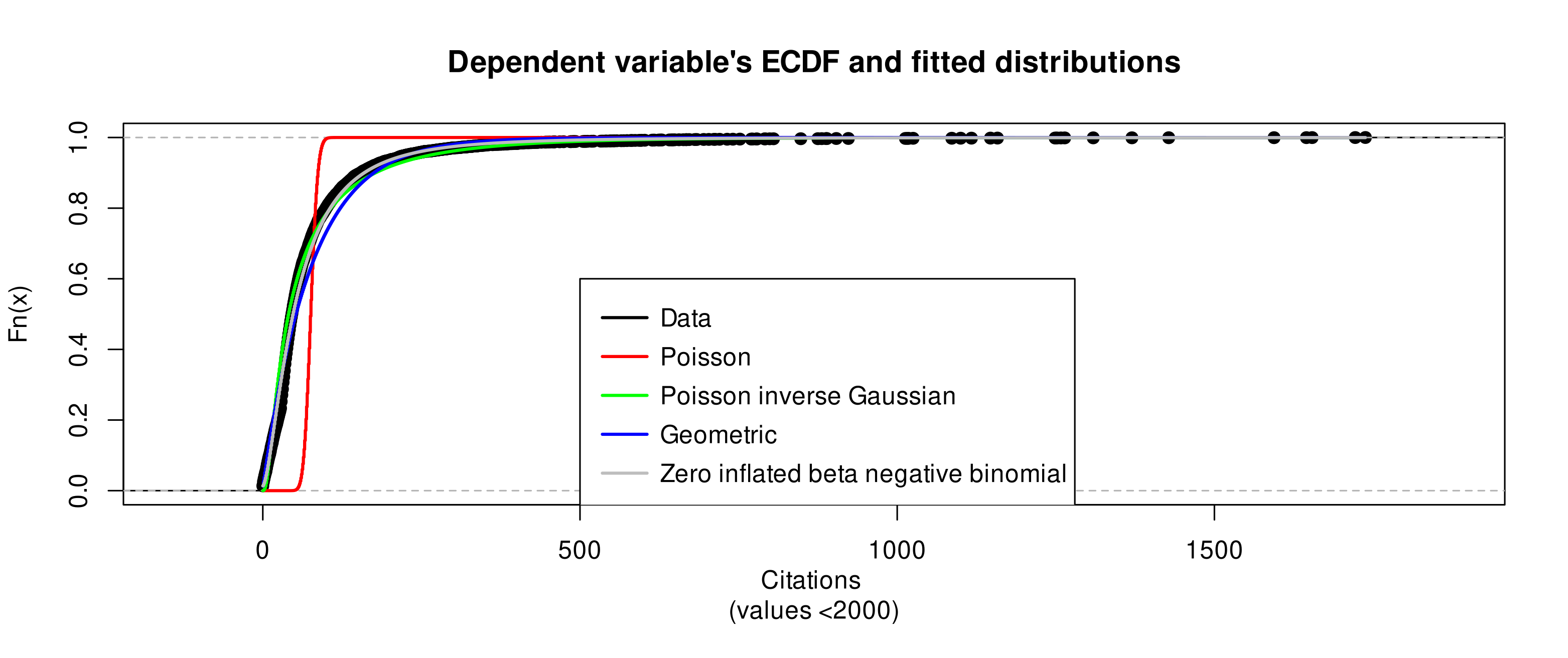}
\caption{The data's ECDF plot and the ECDFs of four fitted distributions.}
\label{fig5}
\end{figure}

Generalized Additive Models for Location, Scale and Shape (GAMLSS) is the most optimal and flexible approach for modeling these data \cite{Stasinopoulos2017}. GAMLSS allows fitting several count distributions to the marginal distribution and compare their goodness of fit via the Generalized Akaike Information Criterion (GAIC). Table \ref{table10} shows the GAIC results of the different tested distributions. The results indicate that the Zero Inflated Beta Negative Binomial (ZIBNB) and the Zero Adjusted (Hurdle) Beta Negative Binomial (ZABNB) distributions gave the best fit. Figure \ref{fig5} shows the empirical cumulative distribution function (ECDF) plots of the data and four adjusted distributions.

\begin{table}[h!]
\centering
\caption{Results of the GAMLS fit procedure. Distributions are sorted in ascending order according to their GAIC values.}

\label{table10}
\begin{tabular}{lc}
\hline
Family	&	GAIC	\\
\hline
ZABNB	&	85642.27	\\
ZIBNB	&	85642.27	\\
BNB	&	85808.65	\\
WARING	&	86225.27	\\
ZISICHEL	&	86445.57	\\
ZASICHEL	&	86445.57	\\
GPO	&	86595.41	\\
ZIPIG	&	86638.7	\\
ZAPIG	&	86638.7	\\
ZINBI	&	86663.48	\\
ZANBI	&	86663.48	\\
NBII	&	86664.83	\\
NBI	&	86664.83	\\
ZINBF	&	86665.48	\\
ZINBF.1	&	86665.48	\\
SICHEL	&	86666.83	\\
NBF	&	86666.83	\\
SI	&	86666.83	\\
DEL	&	86666.83	\\
GEOM	&	86667.32	\\
GEOMo	&	86667.32	\\
PIG	&	87440.66	\\
DPO	&	88590.02	\\
ZALG	&	94111.61	\\
ZAZIPF	&	102051.72	\\
YULE	&	124509.51	\\
ZAP	&	690181.29	\\
ZIP2	&	690181.29	\\
ZIP	&	690181.29	\\
PO	&	705947.67	\\
\hline
\end{tabular}
\end{table}

\subsubsection{Statistical analyses}
As shown above, the four-parameters Zero inflated beta negative binomial (`ZIBNB`) gave the best fit. Hence, the data were modelled with this distribution. 
The first two parameters of the `ZIBNB` are $\mu$ and $\sigma$ and they represent the distributions' location and scale. Recall that GAMLSS enables to model location (e.g. mean), scale (e.g. SD) and shape (i.e. skewness and kurtosis). For simplicity, though, only the location is modelled.
For the case of numeric covariates, besides linear modeling, GAMLSS allows modeling covariates via smothers (e.g. penalized B-splines, monotone P-splines, loess curves). Also, via the package `gamlss.util` it is possible to use neural networks, decision tress, and others (see pages 24 to 25 in \cite{Stasinopoulos2017}). The results of the location modeling are shown in Table \ref{table11} (these results are ranked according to their absolute $t$-values). 

\begin{table}[h!]
\centering
\caption{Ranking of the variables based on their absolute  values. Values for intercepts are not shown.}
\label{table11}
\begin{tabular}{lcccc}
\hline
	&	                 Estimate	&	 Std.Error	   &$t$-value	&   $p$-value	\\
\hline
scihubN	&	          9.884e-03 &	3.742e-04	&	2.641e+01	&	1.525e-147	\\
Total.of.figures.y	& 3.861e-02	&	1.630e-03	&	2.368e+01	&	5.582e-120	\\
IF	&      	          3.663e-02	&	1.920e-03	&	1.907e+01   &	2.070e-79	\\
Hindex1	&	          8.036e-03	&	5.683e-04	&	1.413e+01	&	7.434e-45	\\
Chaperon.Degree	&	  1.787e-01	&	1.594e-02	&	1.121e+01	&	5.716e-29	\\
AuthorsPerArticle &	  8.552e-03	&	7.791e-04	&	1.097e+01	&	7.725e-28	\\
HindexN	&	          1.345e-03	&	2.087e-04	&	6.443e+00	&	1.236e-10	\\
Total.of.tables.y &	  5.781e-03	&	2.937e-03	&	1.968e+00	&	4.905e-02	\\
NatureIndex_A2	&	 -5.845e-05	&	3.312e-05	&  -1.764e+00	&	7.769e-02	\\
NatureIndex_A1	&	 -4.364e-05	&	3.276e-05	&  -1.332e+00	&	1.828e-01	\\
GDPpc_A1	&	     -9.946e-07	&	3.689e-06	&  -2.696e-01	&	7.874e-01	\\
TitleLength	&	     -4.765e-04	&	2.331e-03	&  -2.044e-01	&	8.380e-01	\\
GDPpc_A2	&	      5.401e-07	&	3.577e-06	&   1.509e-01	&	8.799e-01	\\
Pages	&	         -8.889e-07	&	2.122e-04	&  -4.188e-03	&	9.966e-01	\\
\hline

\end{tabular}
\end{table}

\subsubsection{Final Model}
A forward and backward stepwise variable selection procedure applied to a model with all variables, suggested the following model: 

\begin{dmath}
C_i = \alpha+ \beta_1 \times SciHub_i + 
\beta_2 \times APA_i + \beta_3 \times TL_i + \beta_4 \times HIN_i + \beta_5 \times HI1_i +\beta_6 \times CE_i\\ + \beta_7 \times TG_i + \beta_8 \times TT_i + \beta_9 \times IF_i + \beta_{10} \times GDPpc_i + \beta_{11} \times NI_i + u_i     
\end{dmath}

where, $C_i$ is the number of citations the paper $i$ has received; $SciHub$ is the number of times the paper $i$ was downloaded through SciHub; $APA$ is the number of authors per article; $TL$ is the length of the title of the paper; $HIN$ and $HI1$ are the H-index of the first and last author of each paper; $CE$ is the chaperone  effect; $TG$ and $TT$ are the numbers of graphs and tables included in the paper; $IF$ is the impact factor of the journal where the paper was published;  $GDPpc$ is the GDP per capita of the first author, and  $NI$ is the nature index.    

By removing the variables not included in this model, the AIC went from 82692.05 to 82685.60.

Referring back to Table \ref{table11}, and in order to render the model more parsimonious, it could be argued that if three variables were to be kept, then, in this order, they would be: the number of Sci-hub downloads (ScihubN), and the total number of Figures in the published paper (Total.of.figures.y); 'Citations' being the dependent variable. The impact factor of the publishing journal (IF) could also be considered as another good predictor of citations.

\begin{figure}[h!]
\centering
\includegraphics[width=0.8\textwidth]{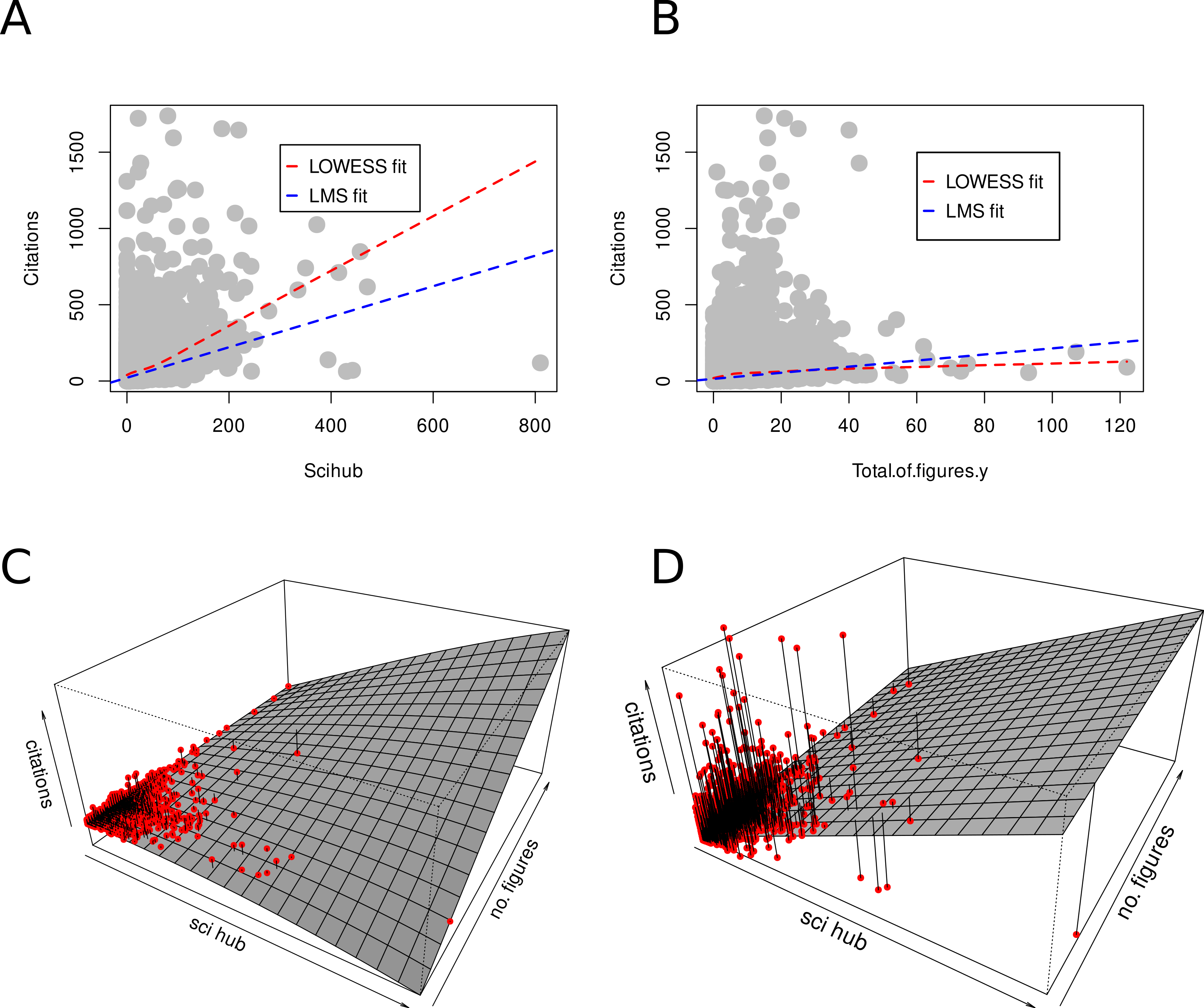}
\caption{Bivariate (A and B) and trivariate associations (C and D) among variables of interest.}
\label{fig6}
\end{figure}

Figure \ref{fig6} displays the results of associations between `Scihub` and `Citations` (A) and between `Total number of figures` and `Citations` (B). Figures \ref{fig6}A and \ref{fig6}B show linear and non-linear fitting lines. The linear fitting is performed via least median of squares (LMS) regression and the non-linear fitting is done via locally weighted scatterplot smoothing (LOWESS). The three-way associations are graphed using ordinary least squares planks (figure \ref{fig6}D) and locally estimated scatterplot smoothing (LOESS).

\section*{Dataset availability}
Our data sets as well as the codes that we developed for the analyses are available in the following public repository.
\url{https://osf.io/8c632/?view_only=19ea965dd02449a0927a3d95d0132a55}.

\end{document}